**The Richest Paradigm You're Not Using: Commercial Videogames at the Intersection of Human-Computer Interaction and Cognitive Science**

Jaap Munneke[1], Jennifer E. Corbett[2,3]

Affiliations

1. Institute for Cognitive and Brain Health, Northeastern University
2. Institute for Medical Engineering & Science, Massachusetts Institute of Technology
3. Computer Science and Artificial Intelligence Lab, Massachusetts Institute of Technology

**Abstract**

Synthesizing from Corbett and Munneke (2025), who demonstrated that questions originating in human-computer interaction (HCI) and game design can be answered through the theoretical toolkit of cognitive science, this perspective argues that commercial videogames represent a largely underutilized research environment at the intersection of these two fields. Cognitive science has long relied on carefully controlled laboratory paradigms to study perception, attention, and executive functioning, raising persistent questions about ecological validity. HCI, by contrast, has spent decades developing methods for studying behavior in rich, complex, interactive environments, but has been less concerned with what that behavior reveals about underlying cognitive mechanisms. Commercial videogames sit precisely at this intersection. They are cognitively demanding by design, motivating by nature, and consistent enough across players to support systematic behavioral comparison. The *affordance structure* of a game does the work that experimental manipulations typically require of the researcher, instantiating cognitive demands that are genuine, sustained, and meaningful to the player. We argue that perception, attention, and executive functioning can be meaningfully studied within commercial games using a minimal observational toolkit of screen recording, eye tracking, and behavioral timing. We propose an affordance-cognition mapping framework as a systematic basis for game selection and research design and offer practical methodological recommendations for researchers wishing to work in this space.

**Commercial Videogames as a Testing Ground**

Every year, cognitive scientists publish thousands of studies on perception, attention, memory, and executive functioning in various academic journals. The vast majority of these studies share a common methodological approach. A participant sits in front of a screen and responds to stimuli that have been carefully stripped of real-world complexity, such as a flanker task (Eriksen & Eriksen, 1974), a spatial cueing paradigm (Posner et al., 1980) or an N-back sequence (Kirchner, 1958). The logic is the same independent of the used paradigm. By reducing the complex world to its essential, barebones variables, it is possible to isolate cognitive mechanisms with high precision. However, this reductionist approach raises a question of whether we are studying cognition as it actually operates in the real world, or are we studying a poor simulation of it? This question is not novel. Nearly 50 years ago, Neisser (1976) asked whether laboratory psychology was producing findings that could ever speak to cognition as it unfolds in the real world. This problem is not simply philosophical in nature as converging evidence shows that performance on certain cognitive tasks fails to predict real-world functioning that relies on the studied cognitive mechanism (Burgess et al., 2006; Chaytor et al., 2006; Chaytor & Schmitter-Edgecombe, 2003; Rosenholtz, 2025). Attention, memory, and executive control as measured in the laboratory may not fully capture how these mechanisms operate when motivation is high, environmental demands are rich, and the consequences of performance are meaningful. This ecological validity question, long present in methodological discussions, points toward a push for expanding the experimental toolkit rather than replacing what already works. Commercial videogames represent a promising, if perhaps unexpected, candidate for bridging this gap.

On any given day, millions of hours of gameplay are logged across the world (Bavelier et al., 2010; Przybylski & Weinstein, 2019). Players navigate three-dimensional environments under time pressure, track multiple moving objects simultaneously, switch rapidly between tasks, form and update spatial representations of vast virtual worlds, and make sequential decisions whose consequences play out across hundreds of subsequent events. Crucially, each of these demands is a direct reflection of deliberate game design choices. From the perspective of a cognitive scientist, this is extraordinary. In a single session of a modern action game such as Call of Duty or Elden Ring, a player may exercise selective attention, sustained vigilance, working memory, inhibitory control, and prospective planning. This all happens within a context that is genuinely engaging to the person gaming. The motivational problem that plagues laboratory research virtually disappears. Participants do not need to be externally motivated into sustained effort. They are, in the most literal sense, playing.

The present perspective argues that commercial videogames, designed purely for entertainment, represent a largely underutilized research environment at the intersection of cognitive science and human-computer interaction (HCI). Specifically, we argue that three foundational cognitive constructs, namely perception, attention, and executive functioning (including working memory), can be meaningfully studied within commercial games using a minimal observational toolkit: screen recording, eye tracking, and behavioral timing (derived from existing in-game metrics or based on precise measurements of screen recordings). This is by no means an argument for abandoning controlled laboratory paradigms, which remain indispensable for mechanistic inquiry. Rather, it is an argument for complementing them with a class of experimental environments in which genuine cognitive demand arises from the structure of the game itself in a context that is inherently motivating, sustained, and meaningful. These are conditions that are at best difficult to replicate in classical laboratory settings

This perspective should be distinguished from a related but separate line of work. A growing and productive research tradition has pursued the gamification of cognitive paradigms (Lumsden et al., 2016). Gamified tasks purposefully redesign experimental tasks to incorporate game-like elements such as points, levels, narratives, and reward structures. This approach has yielded numerous important insights. For example, Seitz and colleagues have demonstrated that embedding perceptual learning within game-like training environments can substantially enhance learning rates and transfer, in part by leveraging the motivational and arousal-related benefits that game mechanics elicit (Deveau et al., 2014; Seitz, 2017). The current paper does share some of the gamification principles in that motivation, engagement, and sustained effort are not noise variables to be controlled away, but meaningful modulators of cognitive performance. However, our focus is different. Rather than designing tasks around cognitive targets, we ask what can be learned from tasks that were never designed with cognitive science in mind at all, but whose design nonetheless structures cognitive demand in ways that can be systematically mapped. This is also where HCI becomes relevant, as researchers have long studied how people behave within complex interactive systems. HCI has spent decades developing practical tools for studying exactly this kind of behavior. However, historically HCI has been less concerned with what that behavior reveals about underlying cognitive architecture. Cognitive science, in turn, has the theoretical tools to make sense of that behavior, but has been slower to incorporate the kind of rich, real-world environments and observational HCI methods. Commercial videogames sit right at this intersection, which is the position this paper takes.

**Two complementary approaches: HCI and Cognitive Science**

HCI is concerned with how people behave when engaging with digital systems in terms of how they navigate interfaces, allocate attention across multiple, competing information sources, make decisions under time pressure, and adapt to novel interaction demands, each of which reflects a deliberate design choice about what the system requires of its user (Carroll, 1997). As a discipline, HCI has developed a robust methodological toolkit for studying exactly these behaviors: interaction logging, behavioral observation, eye tracking, think-aloud protocols, and task performance analyses. Crucially, HCI does not require access to the internal architecture of the systems it studies. Instead, it observes, records, and interprets behavior at the interface.

Cognitive science, by contrast, is concerned with the mechanisms that underlie behavior as representational and computational processes which give rise to constructs such as perception, attention, working memory, and executive functioning. Cognitive science is known for its emphasis on experimental control. Isolating variables, manipulating conditions, and measuring responses with precision has produced an extraordinarily detailed understanding of the human cognitive system, but often at the cost of the very complexity that characterizes cognition in the real world. As noted, the tasks and environments cognitive scientists use to study cognition are, by design, sparse and impoverished relative to the environments in which humans normally operate.

There is a clear asymmetry between the two fields. HCI has the methods to study behavior in rich, complex, naturalistic interactive environments but has historically been less invested in asking what that behavior reveals about underlying cognitive mechanisms (Carroll, 1997). Cognitive science has the theoretical frameworks to interpret behavior in mechanistically meaningful terms but has mostly held on to the experimental control that its methodology depends on. Commercial videogames represent an environment where this asymmetry can be resolved. Games are interactive systems of considerable complexity (think of a game like StarCraft or Horizon Zero Dawn) but they also place sustained, simultaneous demands on perception, attention, memory, and executive functioning in ways that map directly onto cognitive constructs. We argue that this mapping can be made explicit and systematic. Importantly, these constructs can be studied with the tools provided by the field of HCI and interpreted within the theoretical framework cognitive scientists have developed over the last decades. Yet, neither field alone has fully exploited this combination. Our recent work provides an excellent starting point. We (Corbett & Munneke, 2025) used experienced console gamers as participants to investigate why some players invert the y-axis on their controllers. This exemplifies a question that originated in HCI and game design, but whose answer turned out to hinge on individual differences in mental rotation speed and stimulus-response compatibility, two constructs studied within cognitive science. This study required synthesizing literature across HCI, human factors, ergonomics, and cognitive psychology, and serves as a concrete illustration of how questions that originate in one field can be answered by drawing on the theoretical toolkit of another.

It is worth being explicit about what this framing implies. We are not arguing that HCI and cognitive science should merge, nor that game research should replace either field's existing agenda. We are arguing that commercial videogames offer a shared research environment in which the methods of one field and the theory of the other can be put to work together and that doing so stands to benefit both.

**The use of commercial videogames in research**

The argument that commercial videogames can be a productive research environment rests on several converging properties that are difficult to find in any other single context (Sobczyk et al., 2015). Below we outline five properties that we argue reflect a genuine alignment between the structure of gameplay and the demands of cognitive and HCI research.

The first property is *ecological validity*. Most commercial games are not designed to be easy or cognitively undemanding. They are designed to be engaging, which in practical terms means that they impose sustained, varied, and often intense cognitive demands on the player. A player navigating a fast-paced first-person shooter such as Counter Strike must continuously parse a cluttered visual scene, suppress irrelevant distractors, track multiple moving targets, and respond with speed and accuracy, all at the same time. A player managing an empire or a large military in a real-time strategy game (e.g., Age of Empires) must maintain and update multiple representations in working memory, plan several moves ahead, and switch flexibly between competing task demands. These are by no means approximations of cognitive constructs. They are instantiations of them, embedded in a context that is richer and more dynamic than any laboratory task could reasonably replicate (Green & Bavelier, 2012).

The second property is *reproducibility*. An often-heard objection to naturalistic research is that environments vary too much across participants to allow meaningful comparison (Boot et al., 2011). Commercial games largely sidestep this problem. A given level or map presents the same stimuli, the same demands, the same constraints and the same affordance structure to every player who encounters it. In this respect, a commercial game functions as a standardized environment. This environment is far more complex than a laboratory paradigm, but not less consistent across participants. This consistency makes systematic behavioral comparison possible.

The third and perhaps most relevant property for our present argument builds directly on this consistency. The concept of affordances, introduced by Gibson (1979) and developed within HCI by Norman (2013), refers to the action possibilities that an environment offers its user. Game environments are designed to invite specific interactions, and it is this affordance structure that makes them theoretically tractable as research environments. Environments and elements within them draw the player's eye to particular locations, reward particular response patterns, and impose particular sequences of decisions. These affordances emerge from deliberate design choices about what the game demands of its player (Corbett & Munneke, 2025). Crucially, these demands can be mapped onto cognitive constructs without requiring access to the game's underlying code. For example, a researcher does not need to know how the game engine calculates enemy behavior to observe that a player of a fast-paced action game like Overwatch must rapidly detect and respond to threats appearing in the visual periphery. This is a naturalistic example of stimulus-driven exogenous attentional behavior (Posner et al., 1980). The affordance structure of the game does the work of the experimental manipulation.

A fourth and related property is that commercial games are *uncontaminated by experimental intent*. Their demands emerge from the logic of play and from the affordance structure of the game environment, but not from a researcher's hypothesis. This is in sharp contrast to gamified paradigms, which are designed backward from a cognitive target. For instance, a game like StarCraft was not designed to measure cognitive flexibility, but this real-time strategy game imposes exactly the kind of sustained, rapid switching between multiple information sources that cognitive flexibility tasks are designed to capture (Glass et al., 2013). Similarly, a game like The Sims was not designed to study social cognition, yet it places demands on planning, prospective memory, and goal management that map directly onto executive functioning constructs. We argue that the absence of design intent is not a limitation but an asset. It means that any cognitive signal extracted from gameplay reflects something genuine about how the construct operates under naturalistic conditions versus conditions engineered to elicit it.

Finally, commercial games offer a rich and ecologically grounded source of *individual differences* in game related data. Players vary in expertise, strategy, and cognitive/play style. These differences are expressed continuously throughout gameplay rather than in a single response to a probe stimulus. This variance is not 'noise' to be averaged away. It is signal to be interpreted. Individual differences in gameplay behavior can serve as sensitive indices of the cognitive constructs under investigation.

Taken together, these properties make a strong argument for commercial videogames as a potent research environment that is ecologically valid, reproducible, theoretically interpretable, and rich in individual differences data. It is clear that these environments can be used for cognitive and HCI research, but it is not yet understood how to extract valid measures from them or precisely what constructs can be targeted by these metrics. The remainder of this paper addresses both questions.

**What Can Be Measured: An HCI-Informed Observational Toolkit**
Studying cognition inside a commercial videogame presents an immediate practical challenge. The researcher has no access to the game's internal architecture (Sobczyk et al., 2015). There are no event files, no millisecond-precise stimulus logs, no parameter controls. However, the researcher does have everything visible and measurable from the outside including the screen, the player's eyes, and the timing of observable behavior. The question is whether this is enough for research purposes aimed at understanding people's behavior and their interactions with the environment. We argue that it is, and that HCI provides both the methodological precedent and the practical tools to do so.

HCI has never required access to a system's internal architecture to study how people interact with it. Usability research, player experience evaluation, and interaction analysis have all been conducted productively on commercial software using purely observational methods. The tools HCI has developed for this purpose are directly applicable to the study of cognition in gameplay contexts.

*Screen recording* is the most accessible and perhaps most underutilized tool in this context. A continuous recording of the game display captures the full behavioral output of the player over time, including every decision, every movement, and every error. With careful post-hoc event coding, screen recordings yield a rich reconstruction of decision timing, response latencies, error patterns, strategy switches, and spatial navigation choices shaped by the game's affordance structure. In HCI, screen recording and interaction logging have long been used to evaluate how users navigate complex interfaces and where breakdowns in performance occur (Carroll, 1997). In a cognitive science context, the same data stream can be used to index constructs such as inhibitory control (failures to suppress a prepotent response in a game like Overwatch; see Burgess et al., 2006), cognitive flexibility (shifts in strategy following negative feedback in StarCraft II), and prospective memory (whether a player returns to a location or objective after an interruption in an open-world game like The Elder Scrolls or Red Dead Redemption). Importantly, behavioral timing can be extracted from screen recordings with considerable precision through frame-by-frame analysis, going some way toward compensating for the absence of engine-level event logging.

*Behavioral timing* can be derived from frame-by-frame analysis of screen recordings. Many gameplay events present natural choice-reaction opportunities arising from the game's affordance structure, such as a threat appearing in the visual field, a decision point in a navigation task, or a target emerging from behind cover. Response latencies to these events can be extracted through frame analysis, providing ecologically valid reaction time measures without requiring any modification of the game or any awareness on the part of the player that timing is being recorded. In addition to latency, the outcomes of actions can often be scored directly from recordings, such as whether a player successfully responds to a threat, chooses the correct path, or misses a target entirely. This combination of speed and accuracy measures is used in standard laboratory paradigms but can also be captured in a context that the player experiences as entirely natural. The unobtrusiveness of this approach is methodologically useful, as it avoids the demand characteristics that can accompany explicit timing instructions in laboratory settings (Iarygina et al., 2025).

*Eye tracking* is well established in both HCI and cognitive science and is a promising tool for studying cognition during gameplay. In HCI, eye tracking has been used extensively to evaluate interface design, including identifying which elements attract fixations, whether players notice critical information displayed on screen, and whether attentional guidance works as designers intended (Montolio-Vila et al., 2024). In cognitive science, eye tracking provides a continuous, high-resolution window into attentional allocation, scene parsing strategies, and cognitive load. The combination of these two frameworks is particularly powerful in a gameplay context. Where a player looks during gameplay reflects both their attentional priorities and the degree to which the interface successfully guides behavior in terms of whether they fixate on a critical threat, an objective marker, or a peripheral distractor and whether they fixate on the elements the designer intended. Scan path analysis can distinguish top-down, goal-directed gaze patterns from bottom-up, stimulus-driven capture. For example, in studies comparing expert and novice players of StarCraft and League of Legends, experts showed wider gaze distributions and faster saccade velocities than lower-skilled players (Jeong et al., 2022, 2024). Saccade metrics index the speed and efficiency of visual processing under cognitive load and pupil diameter dilates reliably with increasing cognitive demand providing a continuous, unobtrusive proxy for cognitive load that requires no behavioral response from the player and no modification of the game environment (Beatty, 1982; Kahneman & Beatty, 1966; van der Wel & van Steenbergen, 2018). These measures are particularly useful in a commercial game context where cognitive load fluctuates continuously and cannot be experimentally manipulated in the conventional sense.

Used together, these three tools can be extremely informative. Screen recording captures what the player does and when, behavioral timing tells us how fast and how accurately the player responds, and eye tracking reveals where attention is directed before and during those actions. Crucially, all three can be collected simultaneously during a single gameplay session, meaning the same moment in the game can be examined from multiple angles at once (Jeong et al., 2022). This is something that is difficult to achieve in standard laboratory settings, where measurements tend to be collected one at a time.

**Cognitive Constructs in Gameplay**

The following three sections examine what this toolkit can reveal about three foundational cognitive constructs within commercial game environments: perception, attention, and executive functioning. For each construct, we consider both what HCI and cognitive science stand to gain, and what the existing empirical literature already demonstrates.

*Perception*: Perception is the process by which sensory information is selected, organized, and interpreted to form a representation of the environment. For both HCI and cognitive science, commercial videogames can be used as powerful tools to understand how people perceive complex dynamic displays. Games present players with rich, rapidly changing visual scenes that demand continuous perceptual processing under time pressure (Bavelier et al., 2010).

From an HCI perspective, perception in games is inseparable from interface design. Game designers make deliberate choices about what should be visible, and where and when this information should be presented. These are affordance decisions with direct cognitive consequences. For example, enemy silhouettes can be contrasted with backgrounds, health bars can be positioned in peripheral vision, and threat indicators can pulse or flash to capture attention. Whether these design-choices succeed in directing perception as intended is an empirical question that eye tracking is well suited to answer. Studies examining gaze behavior during gameplay have shown that players do not always look where designers assume they will, and that perceptual strategies vary considerably across individuals and expertise levels (Mohan et al., 2023). For instance, expert StarCraft players have been shown to distribute their gaze more widely and execute faster saccades than lower-skilled players, covering more of the screen in less time (Jeong et al., 2022). Similar expertise-related differences in fixation duration and gaze distribution have been documented in League of Legends (Jeong et al., 2024). A broader systematic review of eye tracking studies in action video games confirms that gaze behavior during gameplay is sensitive to both task demands and player expertise, though transfer to naturalistic tasks outside the game remains limited (Montolio-Vila et al., 2024). This gap between design intentions and actual perceptual behavior is a central question addressed in HCI research and it has direct implications for cognitive science. If we want to understand how perception operates in dynamic environments, we need to know what players are actually perceiving versus what they are assumed to be perceiving.

From a cognitive science perspective, games have proven to be a productive environment for studying perceptual learning and perceptual capacity. Foundational work by Green and Bavelier (2003) demonstrated that habitual players of action video games outperformed non-players on a range of visual attention tasks. These included the attentional blink paradigm, in which two targets appear in rapid succession and the task is to detect the second target which can be presented before the visual system has finished processing the first, multiple object tracking where the observer must keep track of several independently moving targets among identical distractors, and measures of the spatial distribution of attention that characterize how broadly and accurately a person can monitor information across the visual field. Action game players showed advantages on all three types of tasks, suggesting that their gaming experience had shaped the way they allocate visual attention at a fundamental level. Crucially, a training experiment established causality such that non-players trained on Medal of Honor (a fast-paced first-person shooter presenting cluttered and rapidly changing visual environments) showed marked improvements on these tasks, while those trained on Tetris (a visually simple puzzle game) did not. This landmark finding not only demonstrated perceptual learning from commercial gameplay but showed that this learning transferred broadly to untrained task. These findings run counter to the typical specificity of perceptual learning observed in laboratory training paradigms (Green & Bavelier, 2012). Rather than gameplay experience in general, the fast-paced, visually demanding affordance structure of the action game appears to be the active ingredient. Subsequent work by Green and colleagues (2010) extended these findings to lower-level perceptual capacities, showing that action game players exhibit enhanced visual acuity and contrast sensitivity relative to non-players. Spence and Feng (2010) reviewed a broader body of evidence showing that game-induced perceptual benefits extend to spatial cognition more generally, including mental rotation and the spatial distribution of visual attention. Bediou and colleagues (2023) confirmed in a large-scale meta-analysis that perception is among the cognitive domains most reliably enhanced by action game experience, with cross-sectional effect sizes in the moderate to large range.

What our observational toolkit adds to this picture is the ability to study perceptual behavior as it unfolds during actual gameplay rather than inferring it from performance on separate laboratory tasks administered before and after a training period. Eye tracking during gameplay allows researchers to observe directly how players scan a complex scene, where they allocate fixations under varying levels of threat and cognitive load, and how perceptual strategies develop with increasing expertise. This is qualitatively different from measuring perceptual capacity in isolation. A player who detects a peripheral threat that the game's design places deliberately in the periphery is not simply demonstrating broad attentional field. Instead, they are doing so while simultaneously managing working memory load, executing motor responses, and monitoring multiple dynamic objects. Eye tracking captures this perceptual behavior in its full cognitive context, something a laboratory attention task by definition cannot do. Screen recording complements this by allowing the researcher to reconstruct exactly what was on screen at the moment of each fixation, providing the stimulus context that a standalone eye tracking dataset would lack.

Taken together, the HCI and cognitive science perspectives on perception converge on the shared conclusion that games are not just environments in which perception happens to be exercised. They are environments in which perception is continuously demanded, shaped by deliberate design choices, and measurable with the tools both fields have already developed. The remaining question is not whether games can tell us something about perception, but how much more we stand to learn once their potential as a research environment is more widely recognized.

*Attention*: Attention is a cognitive construct naturally associated with videogame research. Games place continuous and often extreme demands on attentional resources, requiring players to select relevant information from cluttered environments, sustain attentional focus over extended periods, suppress irrelevant information, and divide attention across multiple sources of information and task demands that arise directly from the game's affordance structure (Sampalo et al., 2023). Both HCI and cognitive science have strong interest in understanding how attention operates in complex interactive environments, and games offer a particularly productive environment for doing so.

From an HCI perspective, attention in games is fundamentally a design issue. Every element of a game interface competes for the player's limited attentional resources, and effective game design depends on understanding how this competition is resolved in practice. Game designers use a range of techniques to guide attention. Motion and animation automatically trigger bottom-up (exogenous) capture, spatial layout and color contrast establish visual priority, and narrative context supports top-down goal-directed search. Whether these techniques work as intended, and under what conditions they break down, are questions that HCI has the tools to answer. Eye tracking studies have shown that attentional capture by interface elements does not always align with designer intent. For example, El-Nasr and Yan (2006) recorded gaze during play across two commercial game genres and found that visual attention patterns were strongly shaped by genre-specific level design, with players in action-adventure games exploring the scene more broadly while first-person shooter players anchored their gaze to the center of the screen. Designers had implicitly structured these patterns through the layout of the environment rather than through explicit attentional cues. High cognitive load conditions can further disrupt attentional guidance in ways that have direct implications for interface design. For instance, the placement of heads-up display elements has been shown to affect how quickly players detect critical information, with consequences not just for user experience but for the cognitive demands the interface imposes (Caroux & Isbister, 2016).

From a cognitive science perspective, the attention literature on commercial games is one of the more developed in the field. Green and Bavelier (2003) established that action game players show superior performance across multiple attentional paradigms, including reduced attentional blink, broader spatial distribution of attention, and enhanced multiple-object tracking capacity. Importantly, training non-players on an action game produced measurable attentional improvements, while training on a non-action control game did not. In successive meta-analyses, Bediou and colleagues (2018, 2023) confirmed that top-down attention is the cognitive domain most robustly and reliably enhanced by action game experience, with effect sizes among the largest observed in the cognitive training literature. Glass and colleagues (2013) extended this to executive aspects of attention, demonstrating that training on the real-time strategy game StarCraft selectively enhanced cognitive flexibility, the ability to rapidly switch between and coordinate multiple information sources, while leaving perceptual speed largely unchanged. This dissociation is theoretically important. It suggests that different game genres utilize different attentional control components, and that genre is a meaningful variable in research design. Sustained attention has received comparatively less systematic study in commercial game contexts, but long gaming sessions constitute a naturalistic vigilance paradigm where participants are genuinely motivated to maintain alertness over extended periods that is virtually impossible to construct in a laboratory.

Our observational toolkit adds the ability to study attentional behavior continuously and in context, rather than through isolated probe tasks. Eye tracking during gameplay captures moment-to-moment attentional allocation across the visual scene, distinguishing between periods of focused, goal-directed search and episodes of bottom-up capture by salient events. Pupil dilation provides a continuous index of attentional load that fluctuates naturally with the demands of the game, allowing researchers to identify which game situations impose the greatest cognitive burden without requiring any interruption to gameplay. Screen recording allows post-hoc identification of the specific game events that precede attentional failures, such as a missed threat, an overlooked objective, or an unnoticed interface element, providing an ecologically grounded account of when and why attention breaks down. Behavioral timing can be used to measure the speed of attentional orienting to game events, yielding reaction time distributions that are directly comparable to those obtained in standard laboratory attention paradigms but embedded in a far richer behavioral context.

The HCI and cognitive science perspectives on attention complement each other in a particularly productive way in this domain. HCI asks whether the interface guides attention effectively and cognitive science asks how much attention the player has available and how efficiently it is deployed. These are different questions that can be answered by the same data, collected with the same tools, during the same gameplay session. This convergence is a clear illustration of what the intersection of these two fields can offer.

*Executive Functioning and Working Memory*: Executive functioning refers to a set of higher-order cognitive processes that manage, control, and coordinate goal-directed behavior. Its core components of working memory, cognitive flexibility, and inhibitory control have been studied extensively within cognitive science. This three-component model proposed by Miyake and colleagues (Miyake et al., 2000) remains the dominant framework for understanding how these capacities relate to one another. For HCI, executive functioning is equally important. Navigating a complex interface, managing multiple task-demands simultaneously, and making rapid decisions under time pressure and uncertainty all draw on the same underlying capacities. Commercial videogames (particularly those that require strategic thinking and real-time resource management) place continuous and severe demands on all three components, making them a particularly productive environment for studying this meta-construct. Basak and colleagues (2008) trained older adults on the real-time strategy game Rise of Nations and found improvements across multiple executive functioning measures including task switching and working memory, while Kühn and colleagues (Kühn et al., 2014) found that training on Super Mario 64 produced structural changes in brain regions associated with spatial navigation, working memory, and strategic planning. Both findings illustrate how the executive demands built into these games, as affordances of the game environment rather than experimentally imposed constraints, can drive meaningful cognitive change.

From a cognitive science perspective, the evidence linking commercial game play to executive functioning is substantial but more nuanced than the evidence for perception and attention. Working memory is perhaps the most well-studied component in this context. Oei and Patterson (2013) demonstrated in a training study using multiple commercial mobile games that spatial working memory specifically improved following training on a spatial memory game, while different genres produced different patterns of transfer consistent with the cognitive demands of each game. Blacker and colleagues (2014) trained participants on Call of Duty: Black Ops and found improvements in visual working memory capacity, pointing to the component-specific nature of game-related executive benefits, with gains tied to the specific attentional affordances the game places on the player rather than reflecting a general executive boost. Glass and colleagues (2013) further demonstrated that training on StarCraft selectively enhanced cognitive flexibility as measured by a broad battery of non-game tasks, while leaving perceptual speed unchanged. This dissociation is important because it maps onto the distinction between fast thinking and fast perception and suggests that the cognitive profile of a game can be characterized with some precision from its design features alone. Dale and Green (2017) reinforced this point in a meta-analysis showing that specific game features like the requirement to manage multiple simultaneous objects and switch between information sources are better predictors of executive function transfer than genre labels alone. This is precisely what an affordance-cognition framework would predict. Inhibitory control has received less systematic attention in commercial game research than working memory and flexibility, but the available evidence is suggestive. By requiring players to suppress responses to irrelevant stimuli while prioritizing targets, action games constitute a naturalistic stop-signal paradigm of sorts. Jakubowska and colleagues (2021) found that actions per minute during StarCraft II play were negatively associated with improvement on a stop-signal task, suggesting that the relationship between game-derived behavioral metrics and inhibitory control is detectable but complex.

Our observational toolkit adds considerably to this picture. Working memory load during gameplay is not directly observable, but it leaves behavioral traces that are. The frequency and pattern of task-switching, the latency between perceiving a game event and responding to it, and the rate of strategic errors are all measurable from screen recordings and behavioral timing, and all reflect the efficiency of executive control in a way that a standalone laboratory task cannot capture in context. For example, in a game like StarCraft II, these traces are continuously generated across an entire match, providing a far richer behavioral record than any single probe task could yield. Eye tracking adds a further layer. Pupil dilation fluctuates with working memory load in real time (Beatty, 1982), and scan path analysis can reveal whether a player is systematically monitoring the information sources that their strategy requires, or whether executive control has broken down in ways that are visible in the gaze record before they are visible in behavior (Hayhoe & Ballard, 2005). Together, these measures allow the researcher to study executive functioning not as an isolated capacity assessed in a single probe task, but as a dynamic process unfolding across hundreds of decisions structured by the game environment itself in a setting where the stakes actually matter to the player.

The convergence of HCI and cognitive science in this domain suggests that both fields are circling around the same questions but have not found a shared language to ask them yet. HCI wants to know how to design interfaces that support rather than overload executive functioning. Cognitive science wants to know how working memory, flexibility, and inhibition operate under genuine cognitive pressure. An affordance-cognition framework offers precisely this shared language where the executive demands built into a game environment are understood as structural properties that are relevant to both the designer and the cognitive scientist. Studied with the tools both fields have developed, commercial videogames offer a context in which both questions can be addressed simultaneously, and where the answers to one are likely to inform the other.

**Towards an Integrated Framework**

In the current perspective we have argued that commercial videogames can be a productive research environment for both HCI and cognitive science. What remains is to develop a clearer understanding of what a genuinely integrated research program would look like. The main difficulty of such a framework is the mapping problem. Given a commercial game, how does a researcher determine what cognitive constructs it is equipped to study and what HCI questions it is able to answer? At present, such a determination is made largely on an ad hoc basis. Researchers select games that seem intuitively relevant to their construct of interest or that have been used in prior work without a clear theoretical rationale for why that game taps the construct of interest. This is by no means a trivial problem. Dale and Green (2017) demonstrated in their meta-analysis that genre labels are poor predictors of cognitive transfer. Instead, they found that what matters is the specific design features of the game. A first-person shooter and a real-time strategy game may both be classified as action games, yet they place fundamentally different demands on perception, working memory, and executive control.

What is needed is an affordance-cognition mapping: a systematic account of how game design features translate into cognitive demands. In a game context, affordances are the interactions the game invites, the decisions it requires of the player, the information it presents, and the responses it rewards or punishes. Mapping these affordances onto cognitive constructs would allow a researcher to approach a commercial game not as a black box, but as a structured cognitive environment with demands that can be characterized in advance of data collection. A game that requires simultaneous tracking of multiple independent agents affords the study of divided attention and working memory capacity. A game that penalizes impulsive responses and rewards delayed action affords the study of inhibitory control. A game that presents the player with a novel environment to be navigated without a map affords the study of spatial memory and cognitive mapping. We (Corbett & Munneke, 2025) have provided the first concrete illustration of this logic by demonstrating that the stimulus-response-effect compatibility structure of $3^{rd}$ person games can be mapped to cognitive constructs like mental rotation and response selection, yielding measurable predictions about individual difference in players' behavior. These mappings are not perfect as games are complex and their demands sometimes interact and overlap. However, games are principled, and they provide a basis for pre-registered research designs that specify in advance which constructs a given game is expected to illuminate and through which observational measures.

Difficulty scaling deserves special mention as a methodological asset that commercial games inherently provide. Most commercial games include adaptive difficulty systems, or at minimum offer discrete difficulty levels that modulate cognitive load in principled ways. From a research perspective, this is equivalent to task demand manipulation that the researcher does not have to engineer. A player progressing through increasingly difficult levels of a game is experiencing a structured increase in cognitive load that can be tracked continuously with our observational toolkit. Pupil dilation will increase, scan paths will change, response latencies will lengthen, and error rates will rise in ways that reflect the cognitive cost of each increment in difficulty. This natural variation in demand is an experimental variable that laboratory paradigms typically have to laboriously construct but that commercial games provide as a design feature.

Individual differences that are often treated as a source of variance to be controlled in laboratory research now become a research asset within this framework. Players vary enormously in expertise, age, cognitive style, and gaming history, and these differences are signals, not noise. A player with high working memory capacity will manage a complex RTS environment differently from one with lower capacity, and those differences will be visible in their gaze patterns, action sequences, and decision latencies. Expertise gradients are particularly valuable. Comparing novice and expert players on the same game and the same observational measures provides a within-game analogue of cross-sectional designs that have been productive in the cognitive training literature, but without requiring a training intervention. The same logic extends to clinical populations. Games that tax specific cognitive systems can in principle serve as ecologically valid assessment environments for populations in whom those systems are compromised, such as patients with frontal lobe damage, individuals with ADHD, or older adults experiencing age-related cognitive decline. The ecological validity and motivational properties of commercial games make them particularly well suited to populations where standard laboratory tasks are poorly tolerated or insufficiently motivating.

It is also important to be clear about what our framework cannot do. Without engine-level access, certain mechanistic questions remain out of reach. The precise timing of stimulus onset, the exact parameters of cognitive load manipulation, and the internal state of the game at any given moment are not directly accessible to the external observer. These constraints mean that the approach described here is better suited to ecological questions like, "How does cognition operate in rich, naturalistic environments?" versus mechanistic ones like, "What are the precise computational processes that underlie a given cognitive operation?" Mechanistic questions are best answered in the laboratory, whereas ecological questions are best answered in our framework. A mature research program can use both, choosing the appropriate environment for the question being asked.

Finally, the development of an integrated research framework between HCI and cognitive science will require methodological standardization. Pre-registration of hypotheses and analysis plans, standardized protocols for screen recording and eye tracking in game contexts, agreed conventions for behavioral coding, and open data sharing could all contribute to a cumulative literature that is currently fragmented across disciplines, methodologies, and game genres. The call for such standardization is not new in either field, but it takes on particular urgency at an intersection where the risk of incompatible methods and incommensurable findings is especially high. Commercial videogames are a shared research environment deserving of a shared methodological infrastructure.

**Methodological Recommendations**

The argument developed in this paper is only as useful as the research it enables. This section offers a set of practical recommendations for researchers wishing to use commercial videogames as a research environment.

1. Game selection should be driven by the affordance-cognition mapping described in the previous section. The cognitive construct or HCI question of interest should determine the game, not the other way around. This selection should be explicitly justified in any resulting publication, with reference to the specific game features that make it appropriate for the question being asked. Genre labels alone are insufficient justification. The observational setup should be standardized within and across studies wherever possible. This includes fixed viewing distance, controlled lighting conditions, consistent eye tracker placement and calibration procedures, and screen recording at a frame rate sufficient for behavioral timing extraction. These are not absolute requirements, but their absence makes cross-study comparison difficult and limits the accumulation of a coherent literature.

2. Within-subject designs are strongly preferable to between-subject designs given the substantial individual variability in gaming experience and cognitive baseline. Where between-subject comparisons are necessary, gaming history should be assessed and reported in detail, as prior experience with a specific game or genre is a meaningful covariate that is routinely underreported in the existing literature.

3. Behavioral coding of screen recordings should follow pre-specified schemes that are developed and piloted before data collection begins. Coders should be blind to participant group or condition, and inter-rater reliability should be standardly reported. The coding scheme itself should be made publicly available to facilitate replication.

4. Pre-registration of hypotheses, game selection rationale, and analysis plans is strongly encouraged. The commercial game literature has historically been vulnerable to flexible analysis and selective reporting. Pre-registration is the most practical safeguard against these problems. Open data sharing, where recordings and eye tracking data can be made available without compromising participant privacy, would further strengthen the cumulative literature.
5. Finally, researchers should be mindful of the ethical and practical considerations that commercial software raises. Institutional licensing, participant consent for recording during gameplay, and the potential for deidentification of behavioral data all require consideration at the design stage rather than as afterthoughts.

**Conclusion**

Commercial videogames are among the most cognitively demanding environments that people voluntarily inhabit. They tax perception, attention, working memory, and executive functioning simultaneously in a context that is motivating, reproducible, and accessible. Nonetheless, they remain largely underutilized as research environments, treated as objects of study rather than tools for studying the mind.

We have argued for a productive path forward that lies at the intersection of HCI and cognitive science. HCI brings a mature observational methodology and a long history of studying behavior in complex interactive systems without requiring access to their internal architecture. Cognitive science brings the theoretical frameworks to interpret that behavior in terms of the constructs that matter most for understanding human cognition. Neither field alone has fully realized what commercial games have to offer. Together, they are well positioned to do so.

Our observational toolkit using screen recording, eye tracking, and behavioral timing is modest in its technical demands but considerable in its reach. It allows the researcher to study perception, attention, memory, and executive functioning as they operate in a genuinely naturalistic context, without sacrificing the systematic rigor that scientific inquiry requires. The constraints imposed by proprietary software are real, but they are clarifying rather than prohibitive. They redirect the researcher toward ecological questions that are, in many respects, more important than the mechanistic ones they preclude.

The games people play at home contain more cognitive science than has yet been extracted from them. This is something we can certainly change with the right questions, the right tools, and a willingness to work across disciplinary boundaries.